# Multilayer $^{10}$B-RPC neutron imaging detector

**L. M. S. Margato**[a,1], **A. Morozov**[a], **A. Blanco**[a], **P. Fonte**[a,b], **L. Lopes**[a], **K. Zeitelhack**[c], **R. Hall-Wilton**[d,e], **C. Höglund**[d,g], **L. Robinson**[d], **S. Schmidt**[d,h], **P. Svensson**[d]

[a] *LIP-Coimbra, Departamento de Física, Universidade de Coimbra, Rua Larga, 3004-516 Coimbra, Portugal*

[b] *Coimbra Polytechnic– ISEC, Coimbra, Portugal*

[c] *Heinz Maier-Leibnitz Zentrum (MLZ), FRM-II, Technische Universität München, D-85748 Garching, Germany*

[d] *European Spallation Source ERIC (ESS), P.O Box 176, SE-221 00 Lund, Sweden*

[e] *ESS and Dipartimento di Fisica "G. Occhialini", Università degli Studi di Milano-Bicocca – Piazza della Scienza 3, 20126 Milano, Italy*

[f] *Linköping University, IFM, Thin Film Physics Division, SE-581 83, Linköping, Sweden*

[g] *Impact Coatings AB, Westmansgatan 29G, SE-582 16 Linköping, Sweden*

[h] *IHI Ionbond AG, Industriestrasse 211, 4600 Olten, Switzerland*

ABSTRACT: Resistive plate chambers (RPC) lined with $^{10}$B$_4$C neutron converters is a promising cost-effective technology for position-sensitive thermal neutron detection capable to outperform $^3$He-based detectors in terms of spatial resolution and timing. However, as for the other types of gaseous detectors with a single layer of $^{10}$B$_4$C at normal beam incidence, the detection efficiency to thermal neutrons of a single-gap $^{10}$B-RPC is only about 6%. Aiming to overcome this limitation, we introduce a multi-layer $^{10}$B-RPCs detector with a stack of ten double-gap hybrid RPCs. A description of the detector design and the results of its characterization performed at the TREFF neutron beamline at the FRM II neutron facility are presented. The results demonstrate that the detection efficiency exceeds 60% for neutrons with a wavelength of 4.7 Å and the spatial resolution (FWHM) is about 0.25 mm and 0.35 mm in the X and Y direction, respectively.

---

[1] E-mail: margato@coimbra.lip.pt

## 1. Introduction

Position-sensitive thermal neutron detectors based on resistive plate chambers (RPC [1]) lined with $^{10}B_4C$ [2] is an emerging technology [5] offering sub-millimeter resolution, sub-nanosecond timing accuracy, and good scalability to large areas at low cost [6,7]. Motivated by the need from neutron scattering facilities, in particular the European Spallation Source (ESS), for large-area PSNDs with cutting edge performance [8,9], a concept of $^{10}$B-RPC position-sensitive neutron detector (PSND) was developed [5] in the frame of SINE2020 project [10].

Sensitivity to thermal neutrons in a $^{10}$B-RPCs detector is achieved through the capture reaction $^{10}B$ (n, α) $^7Li$ in the $^{10}B_4C$ layers [5, 11]. Using a single-gap $^{10}$B-RPC prototype in a hybrid configuration, in which cathodes are metallic [12], we have demonstrated that this detection technology is feasible for position-sensitive thermal neutron detectors: the detector features a wide plateau and the spatial resolution is better than 0.5 mm (FWHM) [12]. However, the thermal neutron detection efficiency obtained with a single converter layer orientated orthogonally to the neutron beam is quite low ($\approx$ 6% at $\lambda = 2.5$ Å) when compared to conventional $^3$He-based detectors. To overcome this limitation, we have proposed several $^{10}$B-RPC detector designs implementing multilayer or inclined layer architectures [5]. Detector designs, implementing the $^{10}B_4C$ neutron converter in multilayers or with the converter layers inclined in relation to the incident neutron beam direction, have also been developed by several groups, mainly based on wire chambers [13-18] but also, e.g., with gas electron multipliers [19].

In this paper we report the results of an experimental study in which one of these solutions to increase the detection efficiency to thermal neutrons is investigated. A $^{10}$B-RPCs detector prototype with 2D position readout is designed using a multilayer architecture in which the RPCs are oriented orthogonally to the incident neutron beam. The prototype is comprised of a stack of ten double-gap RPCs in hybrid configuration [5], with 20 layers of $^{10}B_4C$ in total. The prototype was evaluated at the TREFF neutron beamline (4.7 Å) [20] at the FRM II neutron facility of the Maier-Leibnitz Zentrum (MLZ). It is shown a neutron detection efficiency of $\approx$ 60% and a spatial resolution (FWHM) of 0.25 mm and 0.35 mm in the X and Y direction, respectively.

## 2. Setup and methods

### 2.1. Multilayer $^{10}$B-RPC prototype

The conceptual design of a multilayer $^{10}$B-RPC detector and its working principles are described in our previous paper [5]. The detector prototype, constructed in LIP, is comprised of a stack of ten $^{10}$B-RPC modules (20 layers of $^{10}B_4C$). A thin multi-layer printed circuit board (PCB) with signal pickup strips is inserted between each pair of the neighbouring modules. Figure 1 shows a schematic drawing of the detector layout.

Each $^{10}$B-RPC module consists of a double-gap RPC in hybrid configuration, in which one aluminium cathode plate lined on both sides with $^{10}B_4C$ is shared by two resistive anode plates [5]. Nylon monofilaments with the diameter of 0.35 mm are placed between the anode and the cathode plates to define uniform gas gaps (see Figure 1).

The resistive electrodes (100 mm × 100 mm, 0.5 mm thick) are made of soda lime glass. The plate surfaces on the opposite side of the gas gaps are lined with a thin layer of a resistive ink covering an area of 80 mm × 80 mm. The ink layers, with a surface resistivity of ~ $10^8 \Omega/\square$, are used to evenly distribute the high voltage (HV) across the active area of the anodes. This relatively high resistivity allows to avoid shielding of the signal pickup strips.

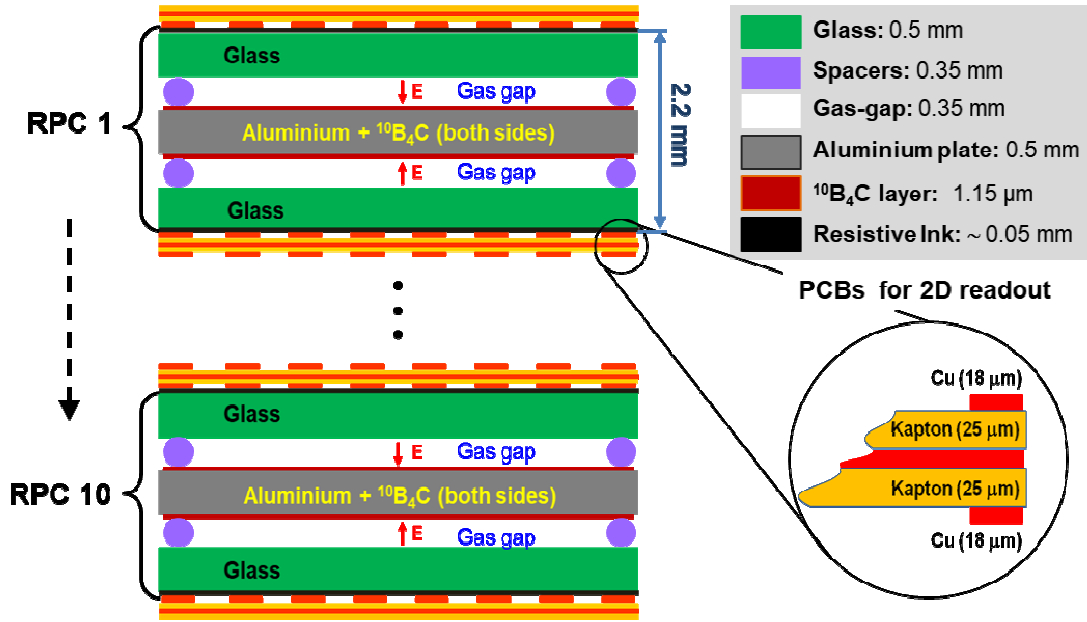

**Figure 1.** Schematic drawing of the multilayer $^{10}$B-RPC detector. The prototype is comprised of ten double-gap hybrid RPC modules, each with an aluminium cathode plate lined on both sides with a 1.15 µm thick layer of $^{10}B_4C$.

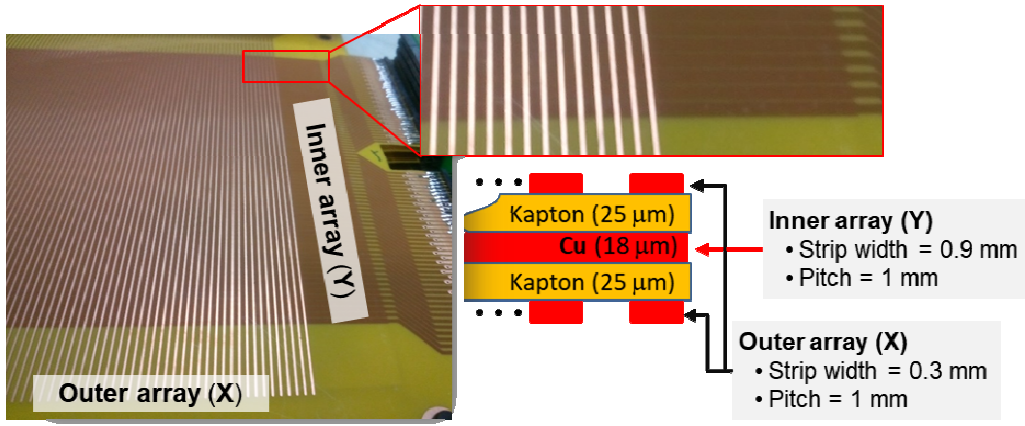

**Figure 2.** PCB with signal pick-up strips for 2D position readout: the strips of the outer and inner arrays are used for readout of X and Y coordinates, respectively.

The aluminium cathode plates (90 mm × 90 mm, 0.5 mm thick), are lined on both sides with a 1.15 µm thick layer of $^{10}B_4C$. The deposition of the $^{10}B_4C$ coatings with a $^{10}B$ enrichment level above 97% has been performed at the Linköping University and the ESS Detector Coatings Workshop in Linköping. During the deposition, the aluminium substrates were kept at temperature bellow 100 ºC minimizing the residual stresses and avoiding their deformation. Further details on the manufacturing of the $^{10}B_4C$ coatings can be found in [2-4].

Flexible multi-layer PCBs are designed to read both X and Y coordinates of a neutron capture event on the anode side. They are inserted between each pair of the neighbouring $^{10}$B-RPC modules. The PCBs feature three arrays of parallel 18 µm thick copper strips with a pitch of 1 mm, separated by 25 µm thick Kapton films (see Figure 2). The inner array strips, designed to read Y coordinate, are oriented orthogonally to the ones of the outer arrays, used to readout X coordinate of the event. The strips are 0.3 mm and 0.9 mm wide in the outer and the inner arrays, respectively. The outer arrays have narrower strips to reduce their screening effect and to have induced signals of a similar magnitude in both arrays. All strips are grounded through 10 MΩ resistors.

To ensure electrical insulation between the strips of the outer arrays and the ink, a grid made of 0.2 mm diameter nylon monofilaments is placed between the RPCs and the PCBs. The RPC anode plates are grounded and the cathode plates are biased with negative voltage.

The stack of ten RPC modules is mounted inside a gas-tight aluminium enclosure. Each module is 2.2 mm thick, and together with the spacers and the PCBs, the stack thickness is 28 mm in total. RPCs are numbered from 1 to 10, with the RPC number one (RPC1) facing the aluminium (1 mm thick) neutron entrance window on the front face of the detector enclosure.

## 2.2. Readout electronics

The RPC cathodes and pickup strips are readout by charge sensitive preamplifiers. The signals from the preamplifiers, assembled on 48-channel front-end electronic boards, are registered by a data acquisition system (DAQ). The DAQ consists of a TRB3 board equipped with two ADC addons based on 40 MHz streaming ADCs (model AD9219from Analog Devices). For a detailed description of the TRB3 platform see [21,22].

The output of each cathode preamplifier is split into two branches. The first branch is fed to a discriminator (model 711 from Phillips) used by the triggering system and the signals in the second branch are digitized by the DAQ. The logic signals from the discriminators are handled by a programmable trigger system [23], which generates the trigger signal for the DAQ and register the cathodes trigger states, which are used to identify the RPC module where the neutron event has occurred. This feature also enables the possibility to determine the neutron capture position along the beam direction, i.e., to measure the neutron time-of-flight (TOF).

Since the triggered RPC module is known from the cathode signals, the strips with the same X index (or, correspondingly, Y index) of all PCBs in the stack are interconnected and readout using the same channel. This approach reduces the number of readout channels by a factor of 20 in comparison with the case of individual readout of each strip.

In total 96 DAQ channels are used for readout of the prototype. Ten channels are allocated for the readout of the cathodes, and the remaining 86 channels are used for the readout of the pickup strips: 43 for the X and 43 for the Y direction, resulting in a readout area of 43 mm × 43 mm.

The counting plateau and detection efficiency measurements are performed using the trigger system by recording the triggering rates of each RPC module. The discriminator threshold is set to -10mV, to avoid triggering with the electronic noise. This threshold voltage is equivalent to a charge of about 60 fC at the preamplifier input.

## 2.3. Experimental system on TREFF neutron beamline

The detector prototype performance is characterized at the TREFF monochromatic neutron beam ($\lambda$ = 4.73 Å) at the FRM II neutron facility of the Maier-Leibnitz Zentrum (MLZ). The detector is positioned on an XY table, with the entrance window oriented normally to the beam (see Figure 3). The irradiated area on the detector is defined by a cadmium slit collimator placed directly in front of the entrance window. The rectangular aperture of the collimator can be controlled with micrometer screws down to the minimum size of 0.1 mm in both vertical and horizontal directions.

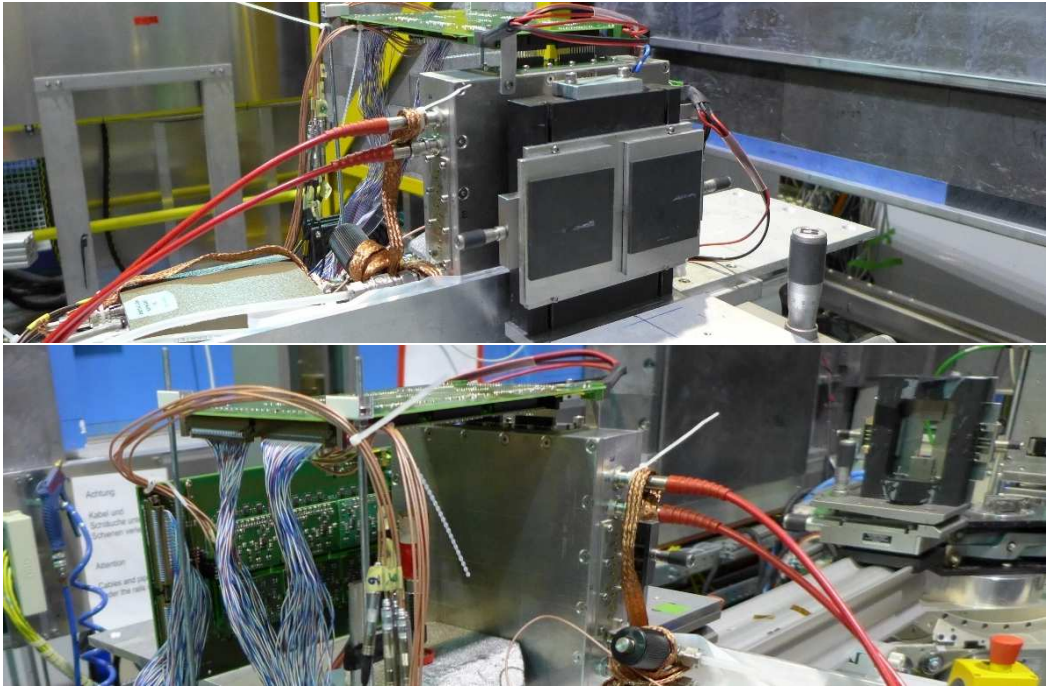

**Figure 3.** Multilayer $^{10}$B-RPC detector installed at the TREFF beamline at FRM II. Top: slit collimator placed in front of the detector. Bottom: detector back view showing elements of the readout electronics.

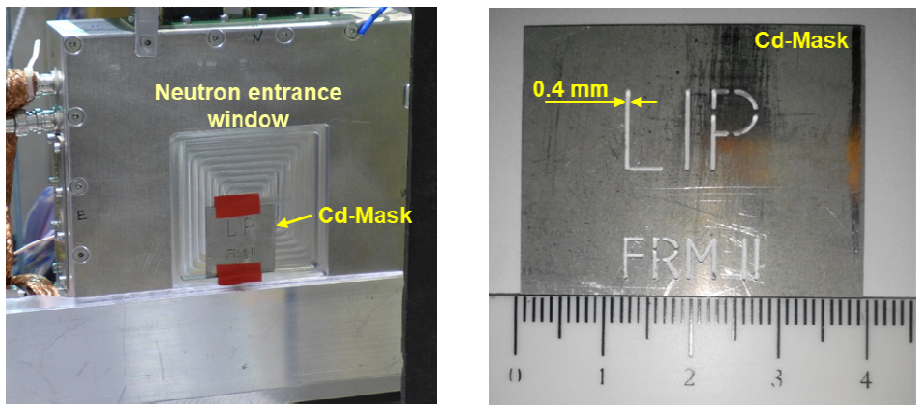

**Figure 4.** Detector with a 1 mm thick cadmium mask (left) and a close-up photograph of the mask (right). The mask is engraved with the words "LIP" and "FRM II". The groves are 0.4 mm wide.

– 5 –

For the imaging experiments the slit collimator is removed and a cadmium mask is placed directly in front of the detector's entrance window (see Figure 4). To measure the beam flux, a calibrated $^3$He-filled proportional counter (97 % detection efficiency for $\lambda = 4.73$ Å) is used as a reference detector.

During the measurements the detector is operated in the avalanche mode, with the working gas ($C_2H_2F_4$) at atmospheric pressure and room temperature (25º C), and maintaining a flow rate of approximately 2 cc/min. The detector is polarized with a negative voltage applied to the cathodes using a high voltage power supply (model 226L from ISEG).

## 2.4. Position reconstruction

The datasets recorded with the DAQ are processed in three steps. First, the signal amplitudes for each cathode and signal pick-up strip are defined for each event: the waveforms are smoothed using 7-point adjacent averaging, and then the maximum amplitude of each waveform and the corresponding baseline are determined. The extracted signal amplitude is set as the difference between these two levels.

In the next step event filtering is performed. An event is rejected if: 1) more than one cathode is triggered (e.g. cosmic ray event); 2) there are signals with amplitude above the saturation level, or 3) there are no strips with signal above the detection threshold in X or Y direction (e.g. neutron capture outside the area covered by the pick-up strips).

In the last step event positions are reconstructed. The XY position is calculated using the center of gravity (COG) algorithm taking into account for each direction only the strip with the strongest signal and the four neighbouring strips on each side. The Z position of the event is assigned according to the triggered cathode.

## 3. Results and discussion

## 3.1. Plateau measurements

The thermal neutron plateau was measured using the trigger system, by recording the overall triggering rate of the $^{10}$B-RPC modules as a function of the HV applied to the RPC cathodes. Figure 5shows the plateau measured for several neutron beam fluxes. The triggering rate is divided by the irradiated area ($2 \times 35$ mm$^2$).

The plateau extends over a range of more than 500 V and the knee appears at ~ 2000 V.Figure 5 also shows the background trigger rate density measured by placing a 6 mm thick boron carbide plate in front of the collimator. The trigger rate in this case is divided by the cathode area ($90 \times 90$ mm$^2$). The main contributions to the background are from the RPC dark counts and from the environmental radiation, including gamma rays originating from the neutron-induced processes in the surrounding materials. The background level is below 1 Hz/cm$^2$ for a larger part of the range of the plateau.

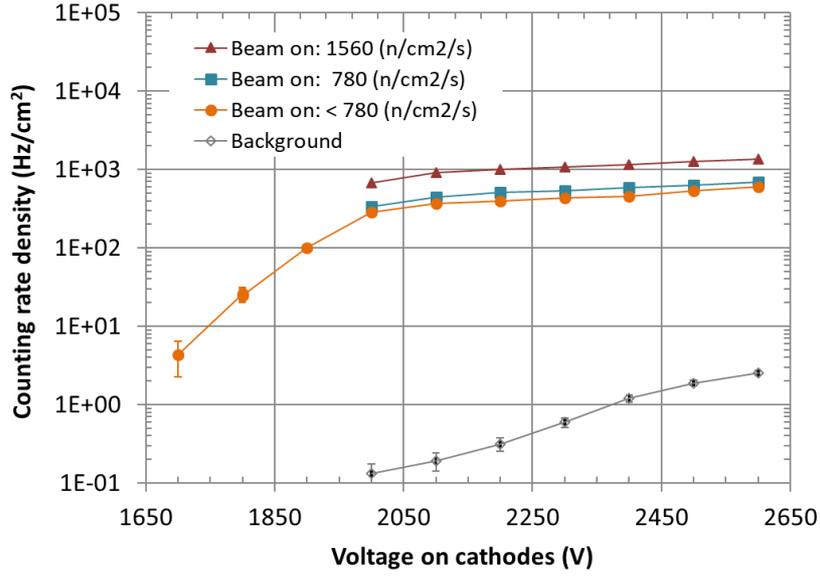

**Figure 5.** Detector counting rate per unit area as a function of the voltage applied to the cathodes for several neutron beam fluxes: triangular, square and round markers correspond to the neutron beam fluxes of $1.56\times10^3$ n/cm$^2$/s, $0.78\times10^3$ n/cm$^2$/s and $< 0.7\times10^3$ n/cm$^2$/s, respectively. Diamond-shaped markers show the background. Error bars show statistical uncertainty and the lines are just to connect the data points and guide the eye.

Two of the three plateau curves presented in Figure 5 are recorded with the known neutron beam fluxes of $0.78\times10^3$ and $1.56\times10^3$ n/s/cm$^2$. All the plateau curves are parallel to each other. The ratio of the count rates given by the two top curves is the same as the ratio of the corresponding neutron fluxes, suggesting the absence of loss of events due to saturation for counting rate densities up to at least 1 kHz/cm$^2$. Note that in this experimental campaign we did not attempt to measure the maximum counting rate density, which should be the target of a dedicated study.

### 3.2. Neutron detection efficiency

The detection efficiency is measured at 2200 V (well within the plateau, see Figure 5) by recording the overall triggering rate of the detector and using an irradiated area of $2 \times 35$ mm$^2$. A neutron event is taken into account when only one of the RPC modules generates a trigger signal. The background is recorded by measuring the triggering rate when the neutron beam is blocked by a 6 mm thick B$_4$C plate. The neutron event rate is calculated by subtracting the background from the overall triggering rate. Finally, the detection efficiency is calculated as the ratio of the neutron detection rate and the neutron beam flux, which is determined using a calibrated $^3$He proportional counter with the nominal detection efficiency of $(97 \pm 1)$ % at 4.73 Å.

The overall detection efficiency of the $^{10}$B-RPC multilayer prototype is $(62.1 \pm 4.5)$ %. The uncertainty is mainly defined by the statistical fluctuations in the recorded triggering rates. The obtained detection efficiency agrees with the Monte Carlo simulation prediction of $\approx 65\%$ for a neutron wavelength of $\lambda = 4.73$ Å [5]. Figure 6 shows the contribution of each RPC module to the overall detection efficiency. The observed exponential decrease in the contributions from the individual RPC module, to the total efficiency is due to attenuation of the neutron beam along its path inside the RPC stack.

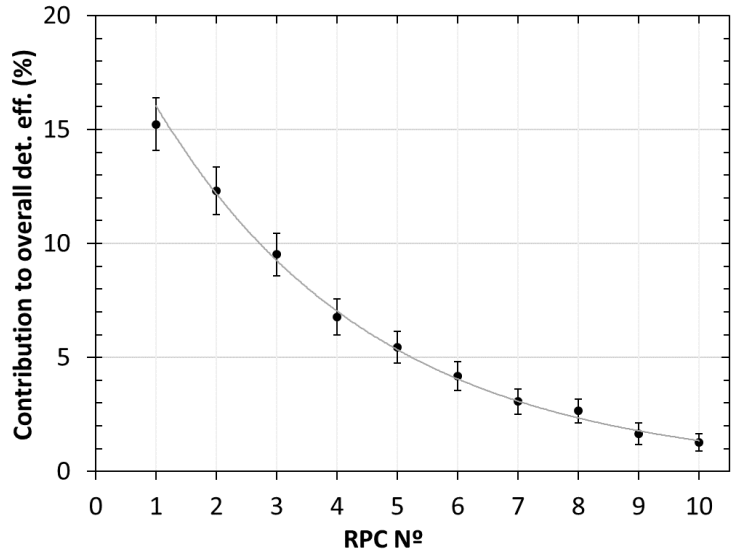

**Figure 6.** Contribution of each $^{10}$B-RPC module to the overall detection efficiency (all values sum up to 62.1%). The cathodes are polarized with -2200 V and the neutron beam flux is $1.56 \times 10^3$ n/cm$^2$/s. Error bars show statistical uncertainty. The line is an exponential fit to the data points.

### 3.3. Imaging performance

To evaluate the spatial resolution for X and Y directions the detector is irradiated with the neutron beam orthogonal to the detector window and collimated by a 0.1 mm wide slit oriented vertically or horizontally. The event positions are reconstructed following the procedure described in section 2.4. The 2D density maps of the reconstructed events position for vertical and horizontal slit orientations are shown in the left-hand side of Figures 7 and 8, respectively. A drop in intensity seen in Figure 7 (left) for Y of about 7 mm is due to a dead electronic channel. The spatial resolution is estimated by performing a gaussian fit to the corresponding profile of the density map projection along the X and Y direction, as shown in the right-hand side of Figures 7 and 8, respectively.

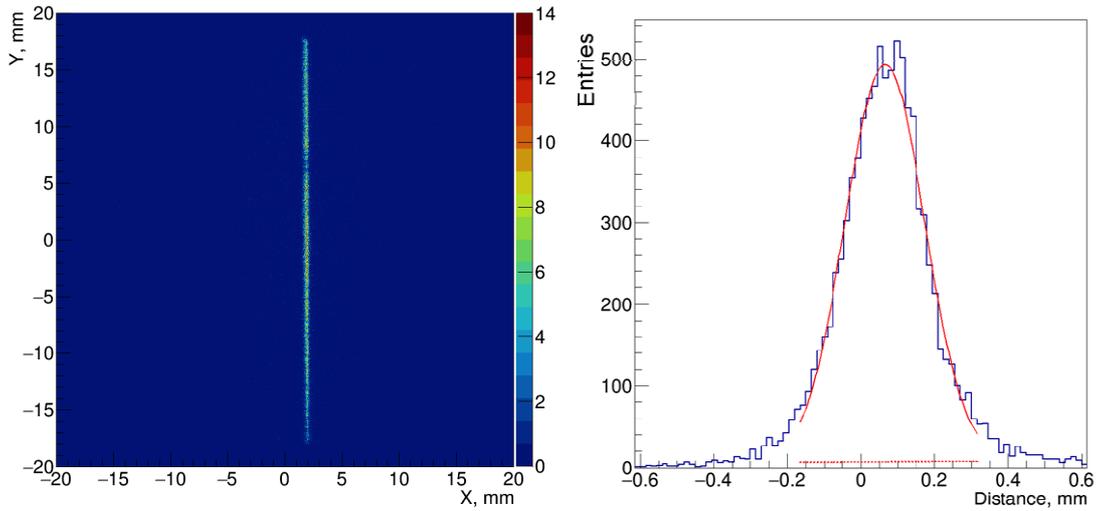

**Figure 7. Left:** 2D density map of the reconstructed event positions (in number of events per bin) obtained for a dataset recorded with the irradiated area of 0.1 mm by 35 mm (vertical slit). **Right**: profile(blue line) of the projection of the density map along the X direction. A gaussian fit of the profile (red line) gives a FWHM of ≈ 0.25 mm. Distance is from the center of the projection extraction box, rotated to be along the slightly tilted slit.

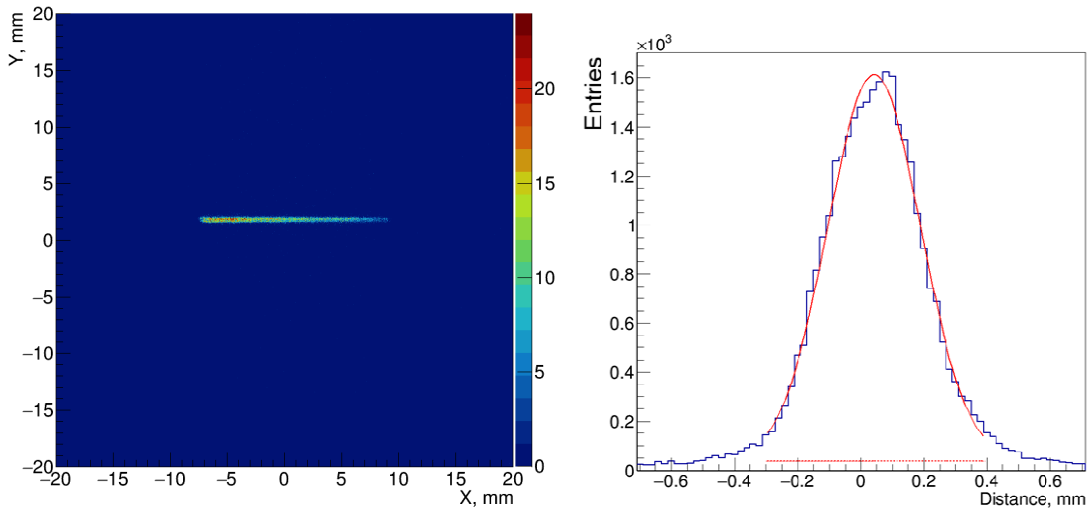

**Figure 8. Left**: 2D density map of the reconstructed event positions (in number of events per bin) obtained for a dataset recorded with the irradiated area of 16 mm by 0.1 mm (horizontal slit). **Right**: profile (blue line) of the projection of the density map along the Y direction. A gaussian fit of the profile (red line) gives a FWHM of≈ 0.35 mm. Distance is from the center of the projection extraction box, rotated to be along the slightly tilted slit.

The results demonstrate a spatial resolution of about 0.25 mm FWHM in the X direction and of about 0.35 mm FWHM in the Y direction. Worse resolution obtained for the Y direction can be explained by assuming that the slit width in this direction was actually larger than the 0.1 mm value set using the micrometer screw of the collimator. This interpretation is based on the fact that the profiles obtained for individual RPCs show a "flat-top" shape as can be seen in Figure 9 for RPC 1 and 2 (not apparent for RPC7 due to low statistics).

It is also observed that the spatial resolution for individual RPCs in the stack essentially does not change with the RPC index as shown in Figure 9. This suggests that neutron scattering does not have a significant effect on the spatial resolution. If that was the case, we would expect worsening of the resolution with the RPC index as the ratio of the detected scattered neutrons and the detected neutrons directly from the beam increases along the stack.

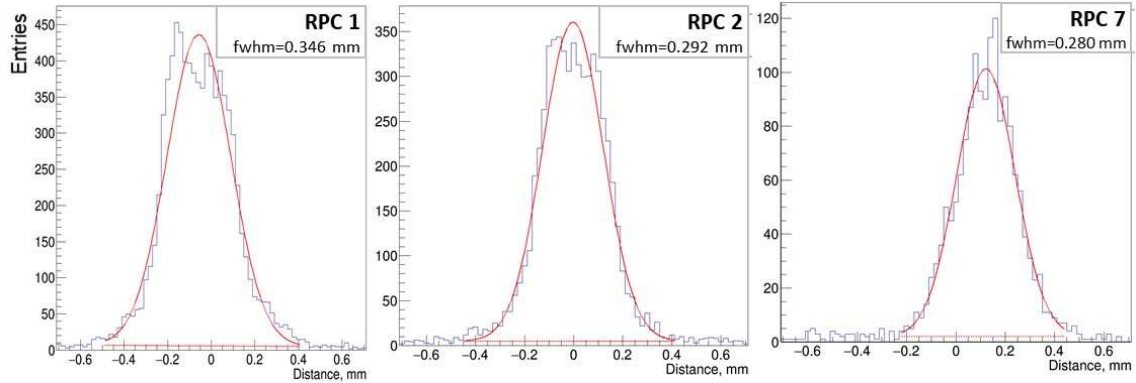

**Figure 9.** Reconstructed neutron beam profile for each individual RPC obtained with the beam collimated slit in the horizontal direction. The distance is from the center of the projection extraction box (exactly the same for all three cases), rotated to be along the slightly tilted slit.

The profiles shown in Figure 7 and 8 are a convolution of the intrinsic resolution with the slit collimator aperture and they are also affected by several factors such as, e.g., the beam divergence, not perfect orthogonality of the incident beam to the RPC stack and small misalignments between the PCBs with the pick-up strips. A systematic shift of the positions of the individual profiles to the right-hand side, shown in Figure 9, can be attributed to non-orthogonality of the neutron beam in respect to the RPC stack: for example, even a small angle of 0.5 degrees results in a significant shift of 0.2 mm over 25 mm along the stack. This systematic shift in the profiles for individual RPC can explains the asymmetry of the overall profile considering all RPCs shown in Figure 8 (right).

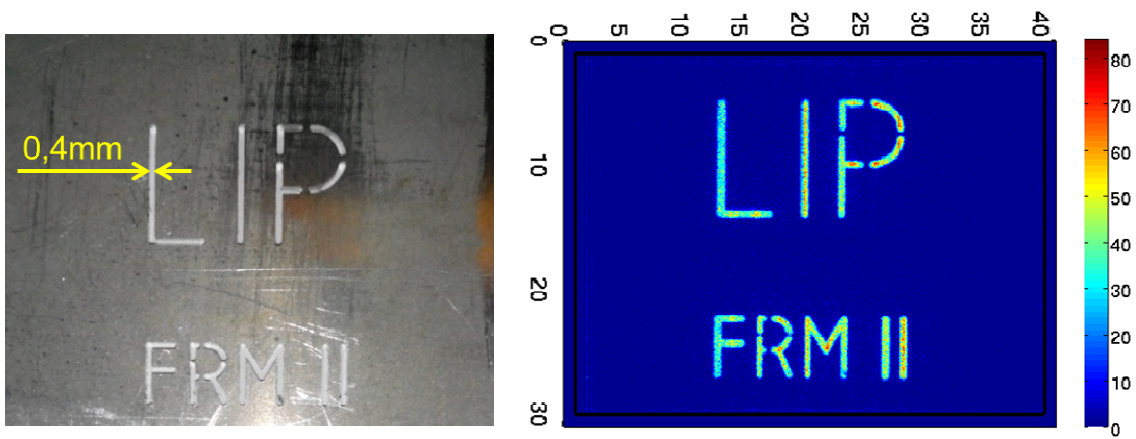

**Figure 10. Left:** Photograph of a 1 mm thick Cd mask. All engraved letters are 0.4 mm wide. **Right**: reconstructed image for a dataset recorded with the mask placed in front the detector entrance window.

Good neutron imaging capability of the detector is also demonstrated in a neutron absorption image of a 1 mm thick cadmium mask. The mask with engraved words "LIP" and "FRM II" (see Figure 4, left)was placed directly in front of the entrance window of the detector. The grooves of the engraved letters are 0.4 mm wide. The neutron event positions were computed by the COG algorithm (see section 2.4). It can be seen in Figure 10 (right) that the reconstructed image reproduces the engraved letters of the mask with high fidelity and even the fine details in the letters P and R are well visible.

## 4. Conclusions and outlook

We report for the first-time operation of a $^{10}$B-RPC detector prototype comprised of a stack of ten double-gap hybrid RPCs. A detection efficiency of (62.1 ± 4.5) % ($\lambda$=4.7 Å) is measured, which is in a good agreement with a prediction of 65% made in our previous simulation study.

The experimental results show a spatial resolution (FWHM) of ≈ 0.25 mm for the X and ≈ 0.35 mm for the Y direction. It is also demonstrated that the spatial resolution of individual RPC modules does not change significantly over the stack.

Very low level of distortions is demonstrated in the experiments with 2D imaging of a cadmium mask with engraved letters. Given the outstanding timing resolution of RPCs [6], the detector can also record the coordinate of a neutron capture event along the beam and therefore provide the time-of-flight information with high precision (e.g., < 1 ns for $^{10}$B$_4$C layers of about 1 µm thick and 1.8 Å neutrons).

The obtained results demonstrate that neutron detection with $^{10}$B-RPC in multilayer architecture is a promising $^3$He alternative technology for construction of PSNDs with high detection efficiency and sub-millimeter spatial resolution. This technology is also suitable for applications in time-resolved neutron imaging (e.g. allowing to follow fast dynamic processes) or for energy-resolved neutron imaging at neutron spallation sources such as the ESS.

**Acknowledgments**

This work was supported by the European Union's Horizon 2020 research and innovation programme under grant agreement No 654000.